\documentclass[aps,prl,superscriptaddress,twocolumn,showpacs]{revtex4}
\usepackage{amsmath}
\usepackage{bm}
\usepackage{graphicx,epsfig}

\renewcommand{\v}[1]{{\bf #1}}
\newcommand{\be}{\begin{equation}}
\newcommand{\ee}{\end{equation}}
\newcommand{\bd}{\begin{displaymath}}
\newcommand{\ed}{\end{displaymath}}
\newcommand{\ba}{\begin{eqnarray}}
\newcommand{\ea}{\end{eqnarray}}
\newcommand{\nn}{\nonumber \\}
\newcommand{\bpm}{\begin{pmatrix}}
\newcommand{\epm}{\end{pmatrix}}

\begin{document}

\title{Skyrmion Dynamics in Multiferroic Insulator}

\author{Ye-Hua Liu}
\affiliation{Zhejiang Institute of Modern Physics and Department of
Physics, \\ Zhejiang University, Hangzhou 310027, People's Republic
of China}
\author{You-Quan Li}
\affiliation{Zhejiang Institute of Modern Physics and Department of
Physics, \\ Zhejiang University, Hangzhou 310027, People's Republic
of China}
\author{Jung Hoon Han}
\email[Electronic address:$~~$]{hanjh@skku.edu}
\affiliation{Department of Physics and BK21 Physics Research
Division, Sungkyunkwan University, Suwon 440-746, Korea}
\affiliation{Asia Pacific Center for Theoretical Physics, Pohang,
Gyeongbuk 790-784, Korea}

\date{\today}

\begin{abstract} Recent discovery of Skyrmion crystal phase in
insulating multiferroic compound Cu$_2$OSeO$_3$ calls for new ways
and ideas to manipulate the Skyrmions in the absence of spin
transfer torque from the conduction electrons. It is shown here that
the position-dependent electric field, pointed along the direction
of the average induced dipole moment of the Skyrmion, can induce the
Hall motion of Skyrmion with its velocity orthogonal to the field
gradient. Finite Gilbert damping produces longitudinal motion. We
find a rich variety of resonance modes excited by a.c. electric
field.
\end{abstract}

\pacs{75.85.+t, 75.70.Kw, 76.50.+g}

\maketitle

Skyrmions are increasingly becoming commonplace sightings among
spiral magnets including the metallic B20
compounds\cite{pfleiderer1,pfleiderer2,tokura,FeGe,MnGe} and most
recently, in a multiferroic insulator Cu$_2$OSeO$_3$\cite{Cu2OSeO3}.
Both species of compounds display similar thickness-dependent phase
diagrams\cite{MnGe,Cu2OSeO3} despite their completely different
electrical properties, highlighting the generality of the Skyrmion
phase in spiral magnets. Along with the ubiquity of Skyrmion matter
comes the challenge of finding means to control and manipulate them,
in a device-oriented manner akin to efforts in spintronics community
to control the domain wall and vortex motion by electrical current.
Spin transfer torque (STT) is a powerful means to induce fast domain
wall motion in metallic magnets\cite{tatara,stiles}. Indeed,
current-driven Skyrmion rotation\cite{torque} and collective
drift\cite{pfleiderer3}, originating from STT, have been
demonstrated in the case of spiral magnets. Theory of
current-induced Skyrmion dynamics has been worked out in Refs.
\cite{zang,rosch}. In insulating compounds such as Cu$_2$OSeO$_3$,
however, the STT-driven mechanism does not work due to the lack of
conduction electrons.

As with other magnetically driven multiferroic
compounds\cite{seki-review}, spiral magnetic order in Cu$_2$OSeO$_3$
is accompanied by finite electric dipole moment. Recent work by Seki
\textit{et al.}\cite{seki} further confirmed the mechanism of
electric dipole moment induction in Cu$_2$OSeO$_3$ to be the
so-called $pd$-hybridization\cite{jia1,jia2,arima}. In short, the
$pd$-hybridization mechanism claims the dipole moment $\v P_{ij}$ for
every oxygen-TM(transition metal) bond proportional to $(\v S_i \cdot
\hat{e}_{ij})^2 \hat{e}_{ij}$ where $i$ and $j$ stand for TM and
oxygen sites, respectively, and $\hat{e}_{ij}$ is the unit vector
connecting them. Carefully summing up the contributions of such terms
over a unit cell consisting of many TM-O bonds, Seki \textit{et al.}
were able to deduce the dipole moment distribution associated with a
given Skyrmionic spin configuration\cite{seki}. It is interesting to
note that the numerical procedure performed by Seki \textit{et al.}
is precisely the coarse-graining procedure which, in the textbook
sense of statistical mechanics, is tantamount to the Ginzburg-Landau
theory of order parameters. Indeed we can show that Seki \textit{et
al.}'s result for the dipole moment distribution is faithfully
reproduced by the assumption that the local dipole moment $\v P_i$ is
related to the local magnetization $\v S_i$ by

\ba \v P_i = \lambda (S^y_i S^z_i , S^z_i S^x_i , S^x_i S^y_i )
\label{eq:pd-coupling}\ea
with some coupling $\lambda$. A similar expression was proposed
earlier in Refs. \cite{penc,han-note} as the GL theory of
Ba$_2$CoGe$_2$O$_7$\cite{BCGO}, another known
$pd$-hybridization-originated multiferroic material with cubic
crystal structure. Each site $i$ corresponds to one cubic unit cell
of Cu$_2$OSeO$_3$ with linear dimension $a\sim 8.9$\AA, and we have
normalized $\v S_i$ to have unit magnitude. The dimension of the
coupling constant is therefore $[\lambda]
=\mathrm{C}\cdot\mathrm{m}$.

Having obtained the proper coupling between dipole moment and the
magnetizaiton vector in Cu$_2$OSeO$_3$  one can readily proceed to
study the spin dynamics by solving Landau-Lifshitz-Gilbert (LLG)
equation. Very small values of Gilbert damping parameter are assumed
in the simulation as we are dealing with an insulating magnet. A
new, critical element in the simulation is the term arising from the
dipolar coupling

\ba H_\mathrm{ME} = -\sum_i \v P_i \cdot \v E_i = -{\lambda \over 2}
\sum_i \v
S_i \bpm 0 & E_i^z & E_i^y \\
E_i^z & 0 & E_i^x \\
E_i^y & E_i^x & 0 \epm \v S_i , \label{eq:H-ME}\ea
where we have used the magneto-electric coupling expression in Eq.
(\ref{eq:pd-coupling}). In essence this is a field-dependent
(voltage-dependent) magnetic anisotropy term. The total Hamiltonian
for spin is given by $H= H_\mathrm{HDM} + H_\mathrm{ME}$, where
$H_\mathrm{HDM}$ consists of the Heisenberg and the
Dzyaloshinskii-Moriya (DM) exchange and a Zeeman field term. Earlier
theoretical studies showed $H_\mathrm{HDM}$ to stabilize the
Skyrmion phase\cite{bogdanov1,bogdanov2,pfleiderer1,han1,han2}.

Two field orientations can be chosen independently in experiments
performed on insulating magnets. First, the direction of magnetic
field $\v B$ determines the plane, orthogonal to $\v B$, in which
Skyrmions form. Second, the electric field $\v E$ can be applied to
couple to the induced dipole moment of the Skyrmion and used as a
``knob" to move it around. Three field directions used in Ref.
\cite{seki} and the induced dipole moment in each case are
classified as (I) $\v B\parallel$ [001], $\v P=0$, (II) $\v
B\parallel$ [110], $\v P\parallel$ [001], and (III) $\v B\parallel$
[111], $\v P\parallel$ [111]. One can rotate the spin axis appearing
in Eq. (\ref{eq:pd-coupling}) accordingly so that the $z$-direction
coincides with the magnetic field orientation in a given setup and
the $x$-direction with the crystallographic $[\overline{1}10]$. In
each of the cases listed above we obtain the magneto-electric
coupling, after the rotation,

\ba H^{(\mathrm{I})}_\mathrm{ME} &=& -{\lambda \over 2} \sum_i E_i
([S^y_i]^2 - [S^x_i]^2) , \nn
H^{(\mathrm{II})}_\mathrm{ME} &=&  -{\lambda \over 2} \sum_i E_i
([S^z_i]^2 - [S^x_i]^2) , \nn
H^{(\mathrm{III})}_\mathrm{ME} &=& -{\lambda \over
{2\sqrt{3}}}\sum_i E_i (3 [S^z_i ]^2 - 1 ) . \label{eq:H-ME2}\ea
In cases (II) and (III) the $\v E$-field is chosen parallel to the
induced dipole moment $\v P$, $\v E_i = E_i \hat{\v P}$, to maximize
the effect of dipolar coupling. In case (I) where there is no net
dipole moment for Skyrmions we chose $\v E
\parallel$ [001] to arrive at a simple magneto-electric coupling
form shown above.

\begin{figure}[ht]
\includegraphics[width=85mm]{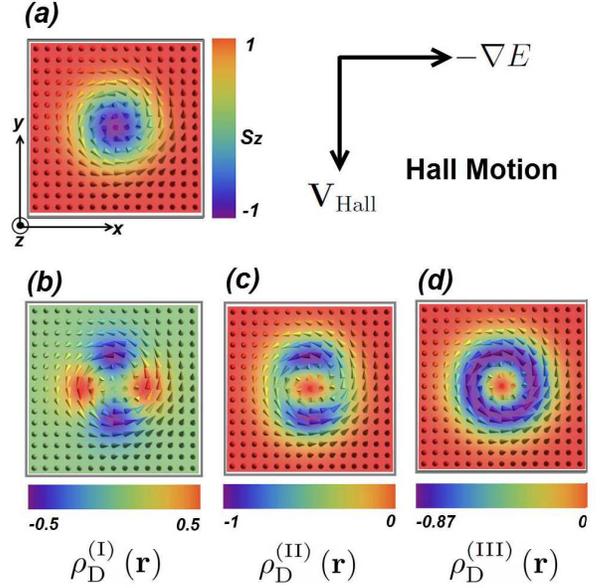}
\caption{(color online) (a) Skyrmion configuration  and (b)-(d) the
corresponding distribution of dipolar charge density for three
magnetic field orientations as in Ref. \onlinecite{seki}. (b) $\v
B\parallel$ [001] (c) $\v B\parallel$ [110] (d) $\v B\parallel$
[111]. For each case, electric field is chosen as $\v E\parallel$
[001], $\v E\parallel$ [001] and $\v E\parallel$ [111],
respectively. See text for the definition of dipolar charge density.
As schematically depicted in (a), the Skyrmion executes a Hall
motion in response to electric field gradient.}\label{fig:rhoeff}
\end{figure}

Suppose now that the $\v E$-field variation is sufficiently slow on
the scale of the lattice constant $a$ to allow the writing of the
continuum energy,

\ba H_\mathrm{ME} = -\lambda { d \over a} \int d^2 \v r E(\v r)
\rho_\mathrm{D} (\v r). \label{eq:continuum-ME}\ea
It is assumed that all variables behave identically along the
thickness direction, of length $d$. The ``dipolar charge" density
$\rho_\mathrm{D} (\v r)$ couples to the electric field $E(\v r)$ in
the same way as the conventional electric charge does to the
potential field in electromagnetism. The analogy is also useful in
thinking about the Skyrmion dynamics under the spatially varying $\v
E$-field as we will show. The continuum form of dipolar charge density
in Eq.~\ref{eq:continuum-ME} is

\ba \rho^{(\mathrm{I})}_\mathrm{D} ({\v r}_i) &=& {1\over 2
a^2 }  ([S^y_i]^2 - [S^x_i]^2)
, \nn
\rho^{(\mathrm{II})}_\mathrm{D} ({\v r}_i) &=& {1\over
2a^2}  ([S^z_i]^2 - [S^x_i]^2
-1) , \nn
\rho^{(\mathrm{III})}_\mathrm{D} ({\v r}_i) &=& {\sqrt{3}\over
2a^2}  ([S^z_i]^2  -1) .
\label{eq:dipolar-charge-density}\ea
Division by the unit cell area
$a^2$ ensures that $\rho_\mathrm{D}(\v r)$ has the dimension of
areal density. Values for the ferromagnetic case, $S^z_i = 1$, is
subtracted in writing down the definition
(\ref{eq:dipolar-charge-density}) in order to isolate the motion of
the Skyrmion \textit{relative to} the ferromagnetic background. Due
to the subtraction, the dipolar charge is no longer equivalent to
the dipole moment of the Skyrmion. The distribution of
dipolar charge density for the Skyrmion spin configuration in the
three cases are plotted in Fig.~\ref{fig:rhoeff}. In case (I) the
total dipolar charge is zero. In cases (II) and (III) the net
dipolar charges are both negative with the relation,
$Q_\mathrm{D}^\mathrm{(II)}/Q_\mathrm{D}^\mathrm{(III)}=\sqrt{3}/2$,
where $Q_\mathrm{D}$, of order unity, is obtained by integrating
$\rho_\mathrm{D} (\v r)$ over the space of one Skyrmion and divide the
result by the number of spins $N_\mathrm{Sk}$ inside the Skyrmion. If the field
variation is slow on the scale of the Skyrmion, then the
point-particle limit is reached by writing $\rho_\mathrm{D} (\v r) =
Q_\mathrm{D} N_\mathrm{Sk} \sum_j \delta (\v r - \v r_j )$ where $\v
r_j$ spans the Skyrmion positions, and identical charge
$Q_\mathrm{D}$ is assumed for all the Skyrmions. We arrive at the
``potential energy" of the collection of Skyrmion particles,

\ba H_\mathrm{ME} =  - \lambda Q_\mathrm{D} N_\mathrm{Sk} {d \over
a} \sum_j E(\v r_j). \label{eq:H-ME3}\ea
A force acting on the Skyrmion will be given as the gradient $\v F_i
= -\bm \nabla_i H_\mathrm{ME}$. Inter-Skyrmion interaction is
ignored.

The response of Skyrmions to a given force, on the other hand, is
that of an electric charge in strong magnetic field, embodied in the
Berry phase action $(-2\pi S \hbar Q_\mathrm{Sk} d/a^3) \sum_j \int dt
(\v r_j \times \dot{\v r}_j ) \cdot \hat{z}$, where $Q_\mathrm{Sk}$
is the quantized Skyrmion charge\cite{zang,stone}, and $S$ is the
size of spin. Equation of motion follows from the combination of the
Berry phase action and Eq. (\ref{eq:H-ME3}),

\ba \v v_j = {\lambda\over 4\pi S\hbar} a^2  N_\mathrm{Sk}
{Q_\mathrm{D}\over Q_\mathrm{Sk}} \hat{z} \times \bm \nabla_j E (\v
r_j ) , \label{eq:Skyrmion-EoM}\ea
where $\v v_j$ is the $j$-th Skyrmion velocity. Typical Hall
velocity can be estimated by replacing $|\bm \nabla E|$ with $\Delta
E/l_\mathrm{Sk}$, where $\Delta E$ is the difference in the field
strength between the left and the right edge of the Skyrmion and
$l_\mathrm{Sk}$ is its diameter. Taking $a^2 N_\mathrm{Sk}\sim
 l_\mathrm{Sk}^2$ we find the velocity

\ba  {\lambda l_\mathrm{Sk}  \over 4 \pi S \hbar } \Delta E  \sim 10^{-6}
\Delta E [\mathrm{m}^2/\mathrm{V}\cdot\mathrm{s}], \ea
which gives the estimated drift velocity of 1 mm/s for the field
strength difference of $10^3$ V/m across the Skyrmion. Experimental
input parameters of $l_\mathrm{Sk}=10^{-7}$m, and $\lambda =
10^{-32}\mathrm{C}\cdot\mathrm{m}$ were taken from Ref. \cite{seki}
in arriving at the estimation, as well as the dipolar and the
Skyrmion charges $Q_\mathrm{D} \approx -1 $ and $Q_\mathrm{Sk} = -1$.
We may estimate the maximum allowed drift velocity by equating the
dipolar energy difference $\lambda \Delta E$ across the Skyrmion to
the exchange energy $J$, also corresponding to the formation energy
of one Skyrmion\cite{han2}.
The maximum expected velocity thus obtained is enormous,
$\sim 10^4$m/s for $J\sim 1\mathrm{meV}$,
implying that with the right engineering
one can achieve rather high Hall velocity of the Skyrmion. In an
encouraging step forward, electric field control of the Skyrmion
lattice orientation in the Cu$_2$OSeO$_3$ crystal was recently
demonstrated\cite{white}.

\begin{figure}[ht]
\includegraphics[width=85mm]{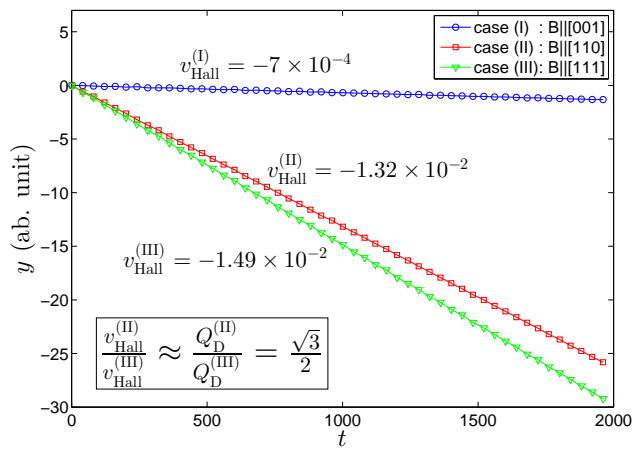}
\caption{(color online) Skyrmion position versus time for cases (I)
through (III) for sinusoidal electric field modulation (see text)
with the Skyrmion center placed at the maximum field gradient
position. The average Hall velocities (in arbitrary units) in cases
(II) and (III) indicated in the figure are approximately
proportional to the respective dipolar charges, in agreement with
Eq. (\ref{eq:Skyrmion-EoM}). A small velocity remains in case (I)
due to imperfect cancelation of forces across the dipolar charge
profile.}\label{fig:speed}
\end{figure}

Results of LLG simulation is discussed next. To start, a sinusoidal
field configuration $E_i = E_0 \sin \left(2\pi x_i /L_x \right)$ is
imposed on a rectangular $L_x \times L_y$ simulation lattice with
$L_{x}$ much larger than the Skyrmion size. In the absence of Gilbert
damping, a single Skyrmion placed in such an environment moved along
the ``equi-potential line" in the $y$-direction as expected from the
guiding-center dynamics of Eq. (\ref{eq:Skyrmion-EoM}). In cases (II)
and (III) where the dipolar charges are nonzero the velocity of the
Skyrmion drift is found to be proportional to their respective
dipolar charges $Q_\mathrm{D}$ as shown in Fig.~\ref{fig:speed}. The
drift velocity decreased continuously as we reduced the field
gradient, obeying the relation (\ref{eq:Skyrmion-EoM}) down to the
zero velocity limit. The dipolar charge is zero in case (I), and
indeed the Skyrmion remains stationary for sufficiently smooth $\v
E$-field gradient. Even for this case, Skyrmions can move for field
gradient modulations taking place on the length scale comparable to
the Skyrmion radius, for the reason that the forces acting on the
positive dipolar charge density blobs (red in Fig.
\ref{fig:rhoeff}(a)) are not completely canceled by those on the
negative dipolar charge density blobs (blue in Fig.
\ref{fig:rhoeff}(a)) for sufficiently rapid variations of the field
strength gradient. A small but non-zero drift velocity ensues, as
shown in Fig. \ref{fig:speed}. Longitudinal motion along the field
gradient begins to develop with finite Gilbert damping, driving the
Skyrmion center to the position of lowest ``potential energy" $E(\v
r)$. For the Skyrmion lattice, imposing a uniform field gradient
across the whole lattice may be too demanding experimentally, unless
the magnetic crystal is cut in the form of a narrow strip the width
of which is comparable to a few Skyrmion radii. In this case we
indeed observe the constant drift of the Skyrmions along the length
of the strip in response to the field gradient across it. The drift
speed is still proportional to the field gradient, but about an order
of magnitude less than that of an isolated Skyrmion under the same
field gradient. We observed the excitation of breathing modes of
Skyrmions when subject to a field gradient, and speculate that such
breathing mode may interfere with the drift motion as the Skyrmions
become closed-packed.

Several movie files are included in the supplementary information.
II.gif and III.gif give Skyrmion motion for $E_i = E_0 \sin
\left(2\pi x_i /L_x \right)$ on $L_x\times L_y = 66\times 66$
lattice for magneto-electric couplings (II) and (III) in Eq.
(\ref{eq:H-ME2}). III-Gilbert.gif gives the same $\v E$-field as
III.gif, with finite Gilbert damping $\alpha = 0.2$. I.gif describes
the case (I) where the average dipolar charge is zero, with a
rapidly varying electric field $E_i = E_0 \sin \left(2\pi x_i
/\lambda_x \right)$ and $\lambda_x$ comparable to the Skyrmion
radius. The case of a narrow strip with the field gradient across is
shown in strip.gif.

Mochizuki's recent simulation\cite{mochizuki} revealed that internal
motion of Skyrmions can be excited with the uniform a.c. magnetic
field. Some of his predictions were confirmed by the recent
microwave measurement\cite{onose}. Here we show that uniform a.c.
\textit{electric field} can also excite several internal modes due
to the magneto-electric coupling. Time-localized electric field pulse was
applied in the LLG simulation and the temporal response $\chi(t)$
was Fourier analyzed, where the response function $\chi(t)$ refers
to $(1/2)\sum_i([S_i^y(t)]^2-[S_i^x(t)]^2)$, $(1/2)\sum_i
([S_i^z(t)]^2-[S_i^x(t)]^2)$, and $(\sqrt{3}/2)\sum_i [S_i^z(t)]^2$ for cases
(I) through (III), respectively.
(In Mochizuki's work, the response
function was the component of total spin along the a.c. magnetic
field direction.)

In case (I), the uniform electric field perturbs the initial
cylindrical symmetry of Skyrmion spin profile so that
$\sum_i([S_i^x(t)]^2-[S_i^y(t)]^2)$ becomes non-zero and the overall
shape becomes elliptical. The axes of the ellipse then rotates
counter-clockwise about the Skyrmion center of mass as illustrated
in supplementary figure, E-mode.gif. There are two additional modes
of higher energies with broken cylindrical symmetry in case (I),
labeled X1 and X2 in Fig. \ref{fig:resonance} and included as X1-mode.gif
and X2-mode.gif in the supplementary. The rotational direction of the
X1-mode is the same as in E-mode, while it is the opposite for
X2-mode.

As in Ref. \cite{mochizuki}, we find sharply defined breathing modes
in cases (II) and (III) at the appropriate resonance frequency
$\omega$, in fact the same frequency at which the a.c. magnetic
field excites the breathing mode. The vertical dashed line in Fig.
\ref{fig:resonance} indicates the common breathing mode
frequency. Movie file B1-mode.gif shows the breathing mode in case
(III). Additional, higher energy B2-mode (B2-mode.gif) was found in
cases (II) and (III), which is the radial mode with one node,
whereas the B1 mode is nodeless.

In addition to the two breathing modes, E-mode and the two X-modes
are excited in case (II) as well due to the partly in-plane nature
of the spin perturbation, $-(\lambda E(t)/2)\sum_i
([S_i^z(t)]^2-[S_i^x(t)]^2)$. In contrast, case (III), where the
perturbation $-(\sqrt{3}\lambda E(t)/2)\sum_i [S_i^z(t) ]^2$ is
purely out-of-plane, one only finds the B-modes. As a result, case
(I) and (III) have no common resonance modes or peaks, while case
(II) has all the peaks (though X1 and X2 peaks are small). Compared
to the magnetic field-induced resonances, a richer variety of modes
are excited by a.c. electric field. In particular, the E-mode has
lower excitation energy than the B-mode and has a sharp resonance
feature, which should make its detection a relatively
straightforward task. Full analytic solution of the excited
modes\cite{tchernyshyov} will be given later.

\begin{figure}[ht]
\includegraphics[width=85mm]{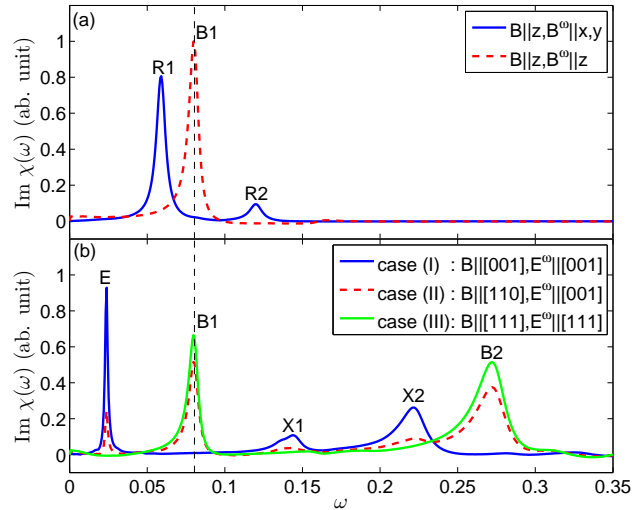}
\caption{(color online) (a) Absorption spectra for a.c. uniform
magnetic field as in Mochizuki's work (reproduced here for
comparison). (b) Absorption spectra for a.c. uniform electric field
in cases (I) through (III). In case (I) where there is no net
dipolar charge we find three low-energy modes E, X1, and X2. For
case (III) where the dipolar charge is finite we find B1 and B2
radial modes excited. Case (II) exhibits all five modes. Detailed
description of each mode is given in the text.}\label{fig:resonance}
\end{figure}

In summary, motivated by the recent discovery of magneto-electric
material Cu$_2$OSeO$_3$ exhibiting Skyrmion lattice phase, we have
outlined the theory of Skyrmion dynamics in such materials. Electric
field gradient is identified as the source of Skyrmion Hall motion.
Several resonant excitations by a.c. electric field are identified.

\acknowledgments J. H. H. is supported by NRF grant (No.
2010-0008529, 2011-0015631). Y. Q. L. is supported by NSFC (Grant No.
11074216). J. H. H. acknowledges earlier collaboration with N.
Nagaosa, Youngbin Tchoe, and J. Zang on a related model and
informative discussion with Y. Tokura.


\begin{thebibliography}{24}

\bibitem{pfleiderer1} S.  M\"{u}hlbauer, B. Binz, F. Jonietz, C.
Pfleiderer, A. Rosch, A. Neubauer, R. Georgii, and P. B\"{o}ni,
Science \textbf{323}, 915 (2009).

\bibitem{pfleiderer2} W. M\"{u}nzer, A. Neubauer, T. Adams, S.
M\"{u}hlbauer, C. Franz, F. Jonietz, R. Georgii, P. B\"{o}ni, B.
Pedersen, M. Schmidt, A. Rosch, and C. Pfleiderer, Phys. Rev. B
\textbf{81}, 041203(R) (2010).

\bibitem{tokura} X. Z. Yu, Y. Onose, N. Kanazawa, J. H. Park,
J. H. Han, Y. Matsui, N. Nagaosa, and Y.  Tokura, Nature
\textbf{465}, 901 (2010).

\bibitem{FeGe} X. Z. Yu, N. Kanazawa, Y. Onose, K. Kimoto,
W. Z. Zhang, S. Ishiwata, Y. Matsui, and Y. Tokura, Nature Mat.
\textbf{10}, 106 (2011).

\bibitem{MnGe} N. Kanazawa, Y. Onose, T. Arima, D. Okuyama, K. Ohoyama, S. Wakimoto,
K. Kakurai, S. Ishiwata, and Y. Tokura, Phys. Rev. Lett.
\textbf{106}, 156603 (2011).

\bibitem{Cu2OSeO3} S. Seki, X. Z. Yu, S. Ishiwata, and Y. Tokura,
\textit{Science} \textbf{336}, 198 (2012).

\bibitem{tatara} G. Tatara, H. Kohno, and J. Shibata, Phys. Rep.
\textbf{468}, 213 (2008).

\bibitem{stiles} D. C. Ralph and M. Stiles, J. Magn. Mag. Mat.
\textbf{320}, 1190 (2008).

\bibitem{torque} F.  Jonietz, S. M\"{u}hlbauer, C. Pfleiderer, A. Neubauer,
W. M\"{u}nzer, A. Bauer, T. Adams, R. Georgii, P. B\"{o}ni, R. A.
Duine, K. Everschor, M. Garst, and A. Rosch, Science \textbf{330},
1648 (2010).

\bibitem{pfleiderer3} T. Schulz, R. Ritz, A. Bauer, M. Halder, M.Wagner,
C. Franz, C. Pfleiderer, K. Everschor, M. Garst, and A. Rosch, Nat.
Phys. \textbf{8}, 301 (2012).

\bibitem{rosch} K. Everschor, M. Garst, R. A. Duine, and A. Rosch,
Phys. Rev. B \textbf{84}, 064401 (2011).

\bibitem{zang} J. Zang, M. Mostovoy, J. H. Han, and N. Nagaosa,
Phys. Rev. Lett. \textbf{107}, 136804 (2011).

\bibitem{seki-review} Y. Tokura and S. Seki, Adv.
Mat. \textbf{21}, 1 (2009).

\bibitem{seki} S. Seki, S. Ishiwata, and Y. Tokura,
arXiv:1206.4404v1 (2012).

\bibitem{jia1} Chenglong Jia, Shigeki Onoda, Naoto Nagaosa, and Jung Hoon
Han, Phys. Rev. B \textbf{74}, 224444 (2006).

\bibitem{jia2} Chenglong Jia, Shigeki Onoda, Naoto Nagaosa, and Jung Hoon
Han, Phys. Rev. B \textbf{76}, 144424 (2007).

\bibitem{arima} Taka-hisa Arima, J. Phys. Soc. Jpn. \textbf{76}, 073702
(2008).

\bibitem{penc} Judit Romh\'{a}nyi,
Mikl\'{o}s Lajk\'{o}, and Karlo Penc, Phys. Rev. B \textbf{84},
224419 (2011).

\bibitem{han-note} J. H. Han, unpublished note.

\bibitem{BCGO} H. Murakawa, Y. Onose, S. Miyahara, N. Furukawa, and
Y. Tokura, Phys. Rev. Lett. \textbf{105}, 137202 (2010).

\bibitem{bogdanov1} A. N.  Bogdanov and D. A. Yablonskii,
Sov. Phys. JETP \textbf{68}, 101 (1989); A. Bogdanov and A. Hubert,
J. Magn. Magn. Mater. \textbf{138}, 255 (1994).

\bibitem{bogdanov2} U. K. Ro{\ss}ler, A. N. Bogdanov, and C. Pfleiderer,
Nature \textbf{442}, 797 (2006).

\bibitem{han1} Su Do Yi, Shigeki Onoda, Naoto Nagaosa, and Jung Hoon
Han, Phys. Rev. B \textbf{80}, 054416 (2009).

\bibitem{han2} Jung Hoon Han, Jiadong  Zang,
Zhihua Yang, Jin-Hong Park, and Naoto Nagaosa, Phys. Rev. B
\textbf{82}, 094429 (2010).

\bibitem{stone} Michael Stone, Phys. Rev. B
\textbf{53}, 16573 (1996).

\bibitem{white} J. S. White, \textit{et al}. arXiv:1208.1146 (2012).

\bibitem{mochizuki} Masahito Mochizuki, Phys. Rev. Lett.
\textbf{108}, 017601 (2012).

\bibitem{tchernyshyov} Olga Petrova and Oleg Tchernyshyov, Phys.
Rev. B \textbf{84}, 214433 (2011); Imam Makhfudz, Benjamin
Kr\"{u}ger, and Oleg Tchernyshyov, arXiv:1208.3123 (2012).

\bibitem{onose} Y. Onose, Y. Okamura, S. Seki, S. Ishiwata, and Y.
Tokura, Phys. Rev. Lett. \textbf{109}, 037603 (2012).

%



\end{thebibliography}
\end{document}